\documentclass[submission, Phys]{SciPost}

\usepackage{amsmath,amssymb}
\usepackage{graphicx}
\usepackage{bm}
\usepackage[utf8]{inputenc}
\usepackage[T1]{fontenc}
\usepackage{xfrac}
\usepackage{mathtools}
\usepackage{braket}
\usepackage{cancel}
\usepackage{xcolor}
\usepackage{float}
\usepackage{hyperref}
\usepackage{subcaption}

\begin{document}

\begin{center}
{\Large \bf Vacancy-Induced Topological Phase Transition \\[4pt]
via Valley Annihilation in an Anisotropic Honeycomb Lattice}

\vspace{10pt}
Anna Hassine$^1$, Amit Goft$^1$, Boris Rotstein$^1$, and Eric Akkermans$^1$

\vspace{4pt}
{\small $^1$~Department of Physics, Technion - Israel Institute of Technology,
Haifa 3200003, Israel}

\vspace{4pt}
{\small \href{mailto:eric@physics.technion.ac.il}{eric@physics.technion.ac.il}}
\end{center}

\begin{abstract}
A single missing atom can drive a topological phase transition in a lattice
that is otherwise trivial for all values of its parameters.
We demonstrate this in a two-dimensional honeycomb lattice with anisotropic
nearest-neighbor hopping ratio $t'/t$.
The pristine lattice is topologically trivial for all $t'/t$ by the
Nielsen-Ninomiya fermion-doubling theorem: Dirac valleys appear in pairs
whose topological charges cancel identically in any bulk invariant.
A single vacancy breaks this cancelation, acting as an internal boundary
with defect winding number $\nu_3=\mp 1$ for $t'/t<2$.
At $t'/t=2$, the two Dirac valleys merge and annihilate; the number of active
pseudospinor degrees of freedom drops from $m=2$ to $m=1$, violating the
condition $d+D+1=2m$ required for a non-trivial winding number.
The winding number collapses to $\nu_3=0$: a topological phase transition
within a \emph{fixed} symmetry class (BDI), driven entirely by a bulk
Lifshitz transition and observable only through the vacancy.
The defect zero mode crosses over from algebraic (${\sim}1/r$) to stronger spatial confinement, with its inverse participation ratio reaching a sharp minimum
at criticality.
Wavefront dislocations in the local density of states provide a direct,
spatially resolved image of $\nu_3$, accessible in graphene and in photonic
and cold-atom analogs.
\end{abstract}

\vspace{6pt}

\vspace{6pt}
\noindent\rule{\linewidth}{0.4pt}

\section{Introduction}
\label{sec:intro}

Topological phases of matter are robust against disorder and deformations,
yet their standard classification assumes translation
invariance~\cite{Altland1997,Kitaev2009,Schnyder2008}.
Real materials unavoidably contain defects, and extending the topological
framework to include them has revealed that point defects can themselves be
the seat of topological invariants~\cite{Teo2010}.
The key quantity is the \emph{Weyl symbol}~\cite{Goft2023,Goft2025_Engineering}
--- the Wigner-Weyl phase-space transform of the Hamiltonian, which plays
the role of the Bloch Hamiltonian in a smoothly inhomogeneous background.
For a chiral system, a non-vanishing defect topological invariant requires
\begin{equation}
    d + D + 1 = 2m ,
    \label{eq:win_condition_chiral}
\end{equation}
where $d$ is the spatial dimension, $D$ the dimension of a sphere encircling
the defect, and $m$ the number of active pseudospinor degrees of freedom
(sublattice, valley, spin).

More precisely, the defect invariant is the winding number $\nu_3$ of the
flattened off-diagonal (chiral) block $q(\bm k,\tilde{\bm r})\in U(m)$ over the
$(d+D)=3$–dimensional manifold that encircles the defect in phase space,
valued in $\pi_{d+D}\big(U(m)\big)=\pi_3\big(U(m)\big)$, which equals
$\mathbb Z$ for $m\ge 2$ and is trivial for $m=1$. A non-vanishing $\nu_3$
therefore requires
\[
2m \;\ge\; d+D+1,
\]
With equality, the marginal case is realized by the honeycomb vacancy below the
transition ($m=2$, $d+D+1=4$). Valley merging reduces $m$ from $2$ to $1$,
violating this bound; since $\pi_3\big(U(1)\big)=0$, the winding necessarily
collapses to $\nu_3=0$.

A fundamental obstruction operates in any periodic bulk: the
Nielsen-Ninomiya theorem~\cite{Nielsen1981a,Nielsen1981b} forces Weyl fermions
to appear in pairs of opposite topological charge, so bulk winding numbers
vanish identically.
The anisotropic honeycomb lattice is a sharp illustration.
It hosts two inequivalent Dirac valleys for $t'/t < 2$ that merge and
annihilate at $t'/t=2$~\cite{Montambaux2009,Montambaux_merging2009}, yet the
lattice remains topologically trivial for all
$t'/t$~\cite{Montambaux2012}.
The Dirac-point merging is a Lifshitz transition of the band structure,
with no topological consequence in the pristine system.
The question we address is whether this changes once a single vacancy is
present.

\paragraph*{Main result.}
We show that the answer is yes.
A single vacancy supports a non-trivial winding number $\nu_3=\mp 1$ for
$t'/t<2$.
At $t'/t=2$, the merging of the two Dirac valleys reduces $m$ from 2 to 1,
invalidating condition~(\ref{eq:win_condition_chiral}), and the winding number
collapses to $\nu_3=0$.
This is a topological phase transition within a \emph{fixed} symmetry class
BDI, invisible in the pristine bulk and observable only through the vacancy.
It is accompanied by a qualitative change in the zero-mode spatial profile
(algebraic to sharp decay) and by a sharp IPR minimum at criticality.
Wavefront dislocations in the local density of states provide a spatially
resolved image of $\nu_3$, accessible in graphene and in analog platforms.

\paragraph*{Relation to prior work and new contributions.}
The Weyl-symbol framework for defect topology, and the existence of a
topologically protected zero mode in the \emph{isotropic} ($t'=t$) honeycomb
lattice with a single vacancy, were established in
Refs.~\cite{Goft2023,abulafia_wavefronts_2023,Goft2025_Engineering,Abulafia2025Potentials}.
The present paper makes the following contributions that go beyond these
results.
\begin{enumerate}
\item \textbf{Topological phase transition.}
We identify and analyse a transition as a function of the continuous parameter
$t'/t$, absent in the isotropic case where the winding number is fixed.
\item \textbf{The $m$-counting mechanism.}
We provide an explicit analytical explanation: the transition is driven by
the reduction of $m$ from 2 to 1 upon valley merging, causing
condition~(\ref{eq:win_condition_chiral}) to fail.
This is qualitatively different from any conventional gap-closing transition.
\item \textbf{IPR fingerprint.}
We identify the IPR minimum as the direct numerical fingerprint of the
transition and confirm its robustness against bond disorder.
\item \textbf{Universality.}
We show that the $m$-counting criterion predicts the transition's existence
and critical point in any bipartite chiral lattice where the valley count
can be continuously reduced, including strained
graphene~\cite{Montambaux2009,Montambaux_merging2009},
Kekul\'e-distorted lattices, and photonic or cold-atom honeycomb analogs.
\end{enumerate}

The paper is organised as follows.
Section~\ref{sec:model} introduces the model.
Section~\ref{sec:below} derives the defect topology below the transition.
Section~\ref{sec:above} analyses the regime above the transition and
identifies the mechanism.
Section~\ref{sec:numerics} presents numerical results.
Section~\ref{sec:discussion} discusses broader implications.

\section{Model and pristine lattice}
\label{sec:model}

\begin{figure}[t]
    \centering
    \includegraphics[width=0.47\textwidth]{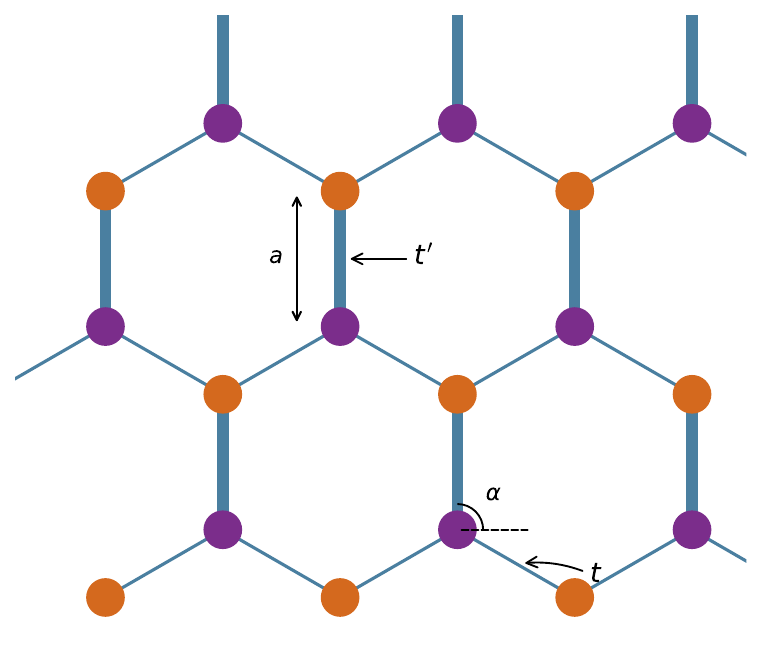}
    \caption{Anisotropic honeycomb lattice with nearest-neighbor hopping
    amplitudes $t$ (thin bonds) and $t'$ (thick bond), parametrised by
    angle $\alpha$. A type-$A$ vacancy (cross) removes one site and
    its three bonds.}
    \label{fig:Pristine_Lattice}
\end{figure}

We consider a nearest-neighbor tight-binding model on a two-dimensional
bipartite lattice with honeycomb geometry
(Fig.~\ref{fig:Pristine_Lattice}).
The lattice is spanned by Bravais vectors
$\boldsymbol{a}_1 = a(\cos\alpha,-\sin\alpha-1)$ and
$\boldsymbol{a}_2 = a(-\cos\alpha,-\sin\alpha-1)$,
where $a$ is the lattice constant and $\alpha\in[0,\pi/2)$ controls the
lattice deformation.
The two sublattices $A$ and $B$ are coupled by anisotropic nearest-neighbor
hopping: one bond per unit cell (along
$\boldsymbol{\gamma}_1 = a(0,1)$) carries amplitude $t'$, while the remaining
two ($\boldsymbol{\gamma}_{2,3}$) carry amplitude $t$.
The Bloch Hamiltonian is
\begin{equation}
    H_0(\boldsymbol{k}) = -\begin{pmatrix} 0 & \tilde{f}(\boldsymbol{k}) \\
    \tilde{f}^*(\boldsymbol{k}) & 0 \end{pmatrix},
    \quad
    \tilde{f}(\boldsymbol{k}) =
    t'e^{-i\boldsymbol{k}\cdot\boldsymbol{\gamma}_1}
    + t\!\!\sum_{j=2,3}\!\!e^{-i\boldsymbol{k}\cdot\boldsymbol{\gamma}_j},
    \label{eq:bloch_hamiltonian}
\end{equation}
with spectrum $\varepsilon(\boldsymbol{k})=\pm|\tilde{f}(\boldsymbol{k})|$.
Chiral symmetry $\{\sigma_z, H_0\}=0$ is exact in the absence of on-site
and next-nearest-neighbor terms.

\begin{figure}[t]
    \centering
    \includegraphics[width=\linewidth]{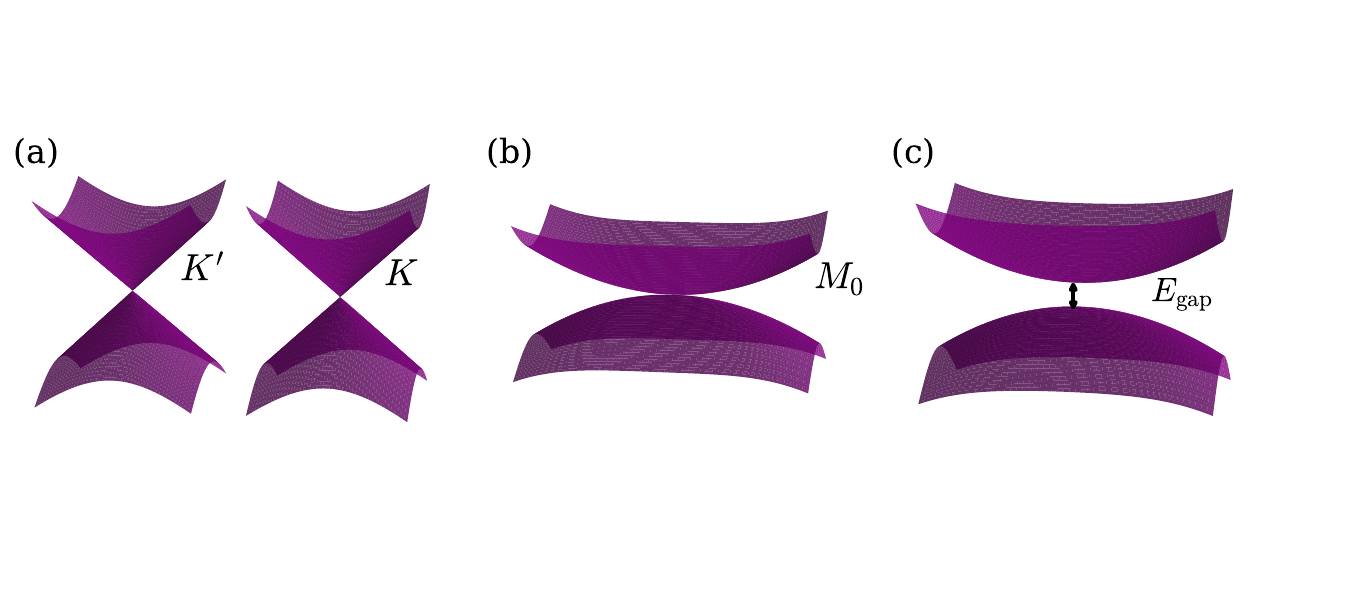}
    \caption{Evolution of the Dirac points with hopping anisotropy $t'/t$.
    (a)~$t'/t<2$: two inequivalent Dirac points (valleys $K$ and $K'$).
    (b)~$t'/t=2$: valleys merge at $\boldsymbol{M}_0$, yielding a semi-Dirac
    dispersion.
    (c)~$t'/t>2$: a spectral gap opens.}
    \label{fig:DP_regime}
\end{figure}

For $t'/t<2$, the spectrum is gapless with two inequivalent Dirac points
$K, K' = \bigl(\pm\tfrac{1}{\cos\alpha}\cos^{-1}(-t'/2t),\,0\bigr)$.
The low-energy Hamiltonian in the valley basis
$\mathcal{B}_{\text{KK'}} = \\ (\ket{AK},\ket{AK'},\ket{BK},\ket{BK'})$ is:
\begin{equation}
    H_0^{\text{below}}(\boldsymbol{k}) =
    -v_x k_x\,\sigma_x\otimes\sigma_z
    -v_y k_y\,\sigma_y\otimes\mathbb{I},
    \label{eq:Hp_below}
\end{equation}
where $v_x = \cos\alpha\sqrt{1-(t'/2t)^2}$,
$v_y = (t'/2t)(\sin\alpha+1)$.
At $t'/t=2$, $v_x\to 0$ and the valleys merge at
$\boldsymbol{M}_0 = (\pi/\cos\alpha, 0)$; for $t'/t>2$ a gap opens
[Fig.~\ref{fig:DP_regime}].
Above the transition, the low-energy Hamiltonian near $\boldsymbol{M}_0$
takes the semi-Dirac (massive) form
\begin{equation}
    H_0^{\text{above}}(\boldsymbol{k}) =
    -(m + u_x k_x^2)\,\sigma_x - v_y k_y\,\sigma_y,
    \label{eq:Hp_above}
\end{equation}
with $m = t'/2t - 1 > 0$, $u_x = \tfrac{1}{2}\cos^2\alpha$.

\paragraph*{Topological triviality of the pristine lattice.}
By the Nielsen-Ninomiya theorem~\cite{Nielsen1981a,Nielsen1981b},
the valleys $K$ and $K'$ carry Berry phases $+1$ and $-1$ that cancel
identically, giving $\nu_0=0$ for all $t'/t$.
The $\sigma_x\otimes\sigma_z$ tensor structure of Eq.~(\ref{eq:Hp_below})
makes this explicit: the two Dirac copies contribute $\pm 1$ to the winding,
summing to zero.
For $t'/t\geq 2$, the mass $m>0$ makes the Fermi-surface phase of
$\tilde{f}(\boldsymbol{k})$ contractible, so $\nu_0=0$ there too.
The Dirac-point merging is a Lifshitz transition, not a topological one.

\section{Defect topology below the transition \texorpdfstring{$t'/t<2$}{t'/t < 2}}
\label{sec:below}

A type-$A$ vacancy at $\boldsymbol{R}_0=\boldsymbol{0}$ is modelled by
removing the three bonds connecting that site to its
neighbors~\cite{Goft2023,abulafia_wavefronts_2023, Abulafia2025Potentials,Kelly1998, Ugeda2010, Ovdat2017, Pereira2006, Pereira2008, Amara2007, Palacios2008, Dutreix2013, Godsil2001, Faccio2010, Yang2018}.
This preserves the chiral symmetry exactly.
For $t'/t<2$, the Weyl symbol of the full Hamiltonian in the valley
basis $\mathcal{B}_{\text{KK'}}$ is~\cite{Goft2023,Goft2025_Engineering}
\begin{align}
H^{\text{below}}(\boldsymbol{k},\tilde{\boldsymbol{r}})
&= v_x k_x\,\sigma_x\otimes\sigma_z
   + v_y k_y\,\sigma_y\otimes\mathbb{I} \nonumber\\
&\quad + \phi_1(\tilde{\boldsymbol{r}})\,\sigma_x\otimes\sigma_x
   + \phi_2(\tilde{\boldsymbol{r}})\,\sigma_x\otimes\sigma_y ,
\label{eq:symbol_below}
\end{align}
where $\phi(\tilde{\boldsymbol{r}}) = \phi_1 + i\phi_2 =
e^{i\tilde\theta}\phi(\tilde r)$ encodes the vacancy perturbation, and
$(\tilde r, \tilde\theta)$ are elliptical coordinates scaled by $(v_x,v_y)$.
The symmetry operators
$T = \mathbb{I}\otimes\sigma_x K$,
$P = \sigma_x\otimes\sigma_x K$,
$C = \sigma_z\otimes\mathbb{I}$
place the system in class BDI.

With $d=2$, $D=1$, and $m=2$ (two active valley pseudospinor degrees of
freedom), condition~(\ref{eq:win_condition_chiral}) gives
$d+D+1 = 4 = 2\times 2 = 2m$: satisfied.
A direct evaluation~\cite{Goft2023,abulafia_wavefronts_2023,Goft2025_Engineering} yields
the defect winding number
\begin{equation}
\nu_3^{\text{below}}
= \frac{1}{2\pi}\int\! d\tilde\theta\;
\frac{1}{\phi_1^2+\phi_2^2}
\begin{vmatrix}
\phi_1 & \phi_2 \\
\partial_{\tilde\theta}\phi_1 & \partial_{\tilde\theta}\phi_2
\end{vmatrix}
= \mp 1 ,
\label{eq:winding_before}
\end{equation}
corresponding to the winding of the phase of $\phi(\tilde{\boldsymbol{r}})$
around the vacancy.
The Atiyah-Singer index
theorem~\cite{Atiyah1968,Eguchi1980, Freed2021, Roe1990, Nakahara1990} then guarantees a single topologically
protected zero-energy mode:
$\mathrm{Index}(\hat{D}) = \dim\ker\hat{D} - \dim\ker\hat{D}^\dagger = 1$
for a type-$A$ vacancy.
The vacancy acts as an \emph{internal boundary}: no physical edge is
required.
For a configuration of $V_A$ ($V_B$) vacancies on sublattice $A$ ($B$),
the theorem gives $\mathrm{Index}(\hat{D}) = V_B - V_A$.


The zero mode is known analytically~\cite{Pereira2006,Abulafia2025Potentials}:
$\psi(\boldsymbol{r})\propto r^{-1}(0,0,e^{i\theta},e^{-i\theta})^T$,
with algebraic $1/r$ decay and weight only on sublattice $B$.
Under the coordinate transformation $(r,\theta)\to(\tilde r,\tilde\theta)$,
the same profile applies to the anisotropic lattice.
Numerical diagonalization on a brickwall lattice reproduces the expected sublattice-polarized, long-ranged zero-mode profile for $t'/t=1$ and $t'/t=1.5$ [Fig.~3].

\begin{figure}[H]
    \centering

    \begin{subfigure}{0.48\linewidth}
        \centering
        \includegraphics[width=\linewidth]{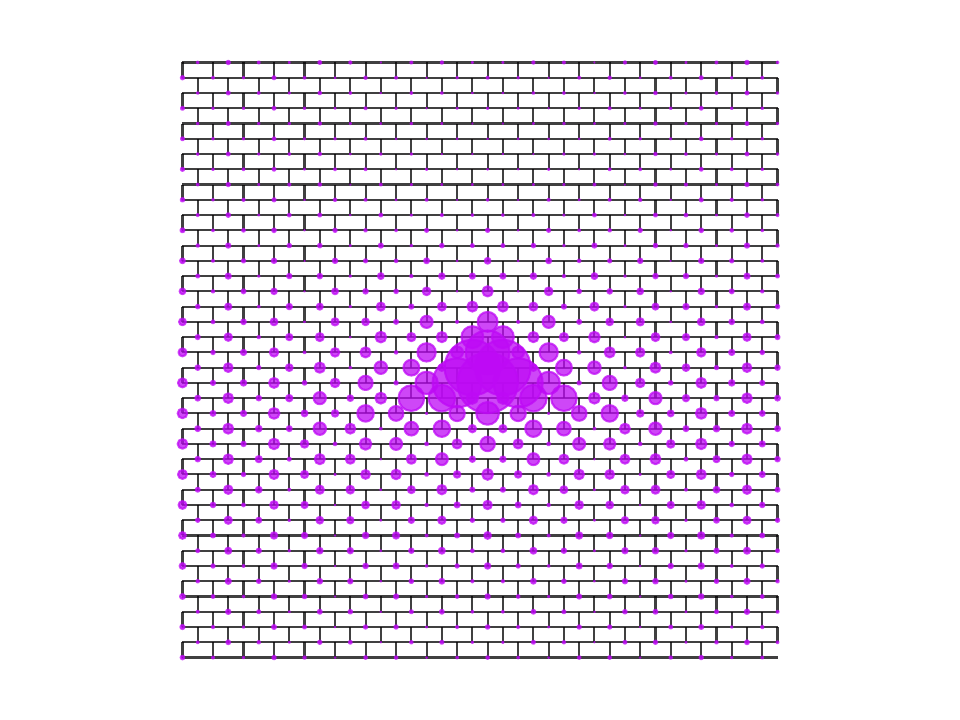}
        \caption{}
        \label{fig:ZM_tp1}
    \end{subfigure}
    \hfill
    \begin{subfigure}{0.48\linewidth}
        \centering
        \includegraphics[width=\linewidth]{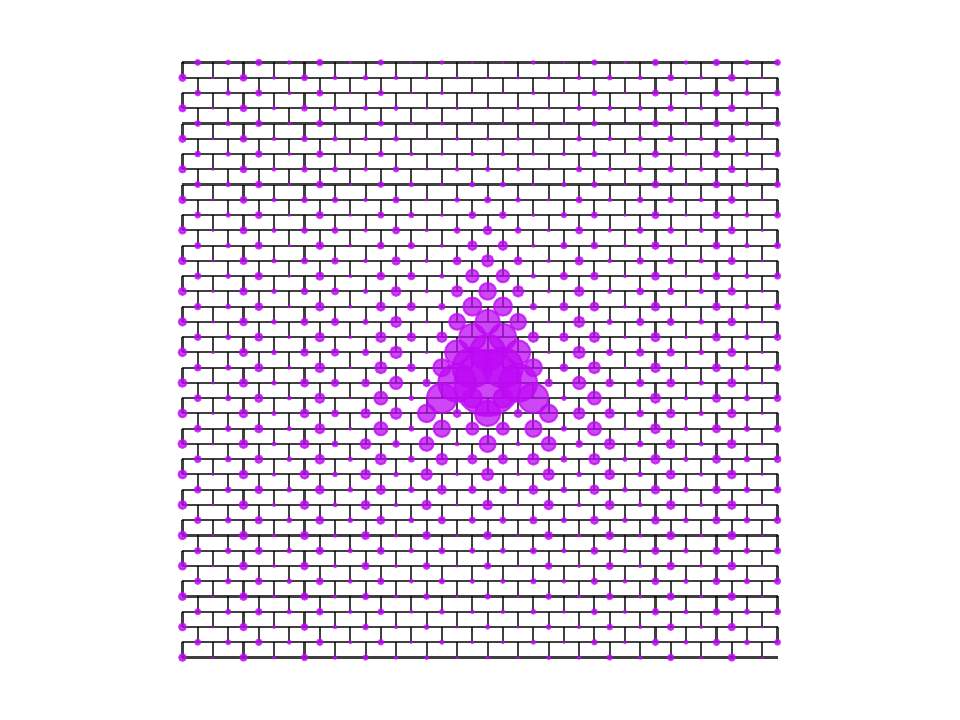}
        \caption{}
        \label{fig:ZM_tp15}
    \end{subfigure}

    \caption{Zero-mode probability density for $t'/t<2$ (periodic boundary
    conditions).
    (a)~$t'/t=1$: long-ranged, weight on sublattice $B$ only.
    (b)~$t'/t=1.5$: mode spreads directionally but remains long-ranged and sublattice polarized.}
    
    \label{fig:ZM_before}
\end{figure}

\section{Defect topology above the transition and the transition mechanism}
\label{sec:above}

\begin{figure}[H]
    \centering

    \begin{subfigure}{0.48\linewidth}
        \centering
        \includegraphics[width=\linewidth]{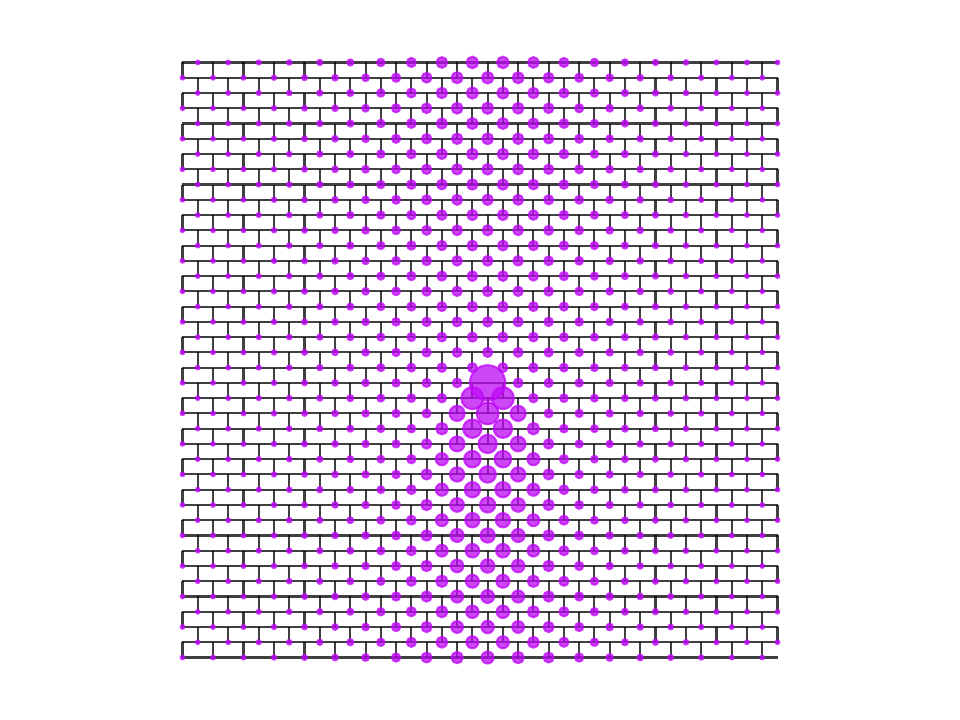}
        \caption{}
        \label{fig:ZM_tp201}
    \end{subfigure}
    \hfill
    \begin{subfigure}{0.48\linewidth}
        \centering
        \includegraphics[width=\linewidth]{
        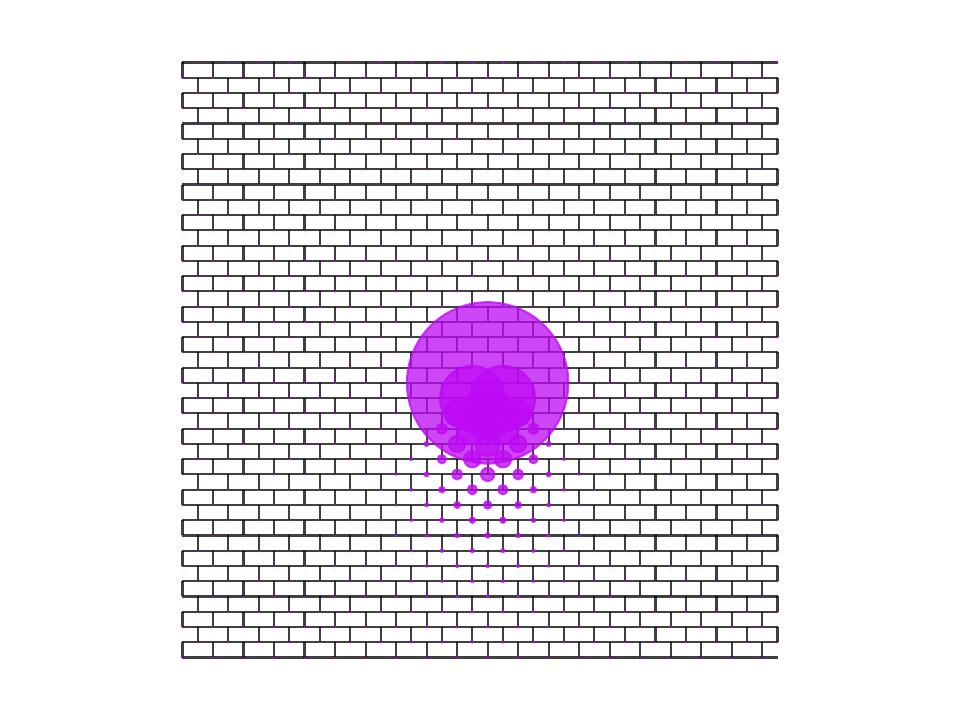}
        \caption{}
        \label{fig:ZM_tp25}
    \end{subfigure}

    \caption{Numerical vacancy zero-mode probability density for $t'/t>2$ under periodic boundary conditions.
    (a)~$t'/t=2.01$: close to the critical point, the mode remains broadly distributed around the vacancy.
    (b)~$t'/t=2.25$: deeper in the gapped phase, the density becomes more strongly confined near the vacancy. Sublattice polarization persists in both panels.}
    \label{fig:ZM_after}
\end{figure}

Before computing the Weyl symbol above the transition, we clarify a
distinction that is inessential below $t'/t=2$ but becomes central at the
transition: the difference between the \emph{existence} of the defect zero
mode and its \emph{winding number}. The existence is fixed by the
Atiyah–Singer index \cite{Atiyah1963, Atiyah1968, Freed2021, Eguchi1980, Roe1990, Nakahara1990} $\mathrm{Index}(\hat D)=V_B-V_A$, which counts only the
sublattice imbalance and is independent of $t'/t$: a single type-$A$ vacancy
supports exactly one chiral zero mode for every $t'/t$, consistent with the
eigenvalue that remains spectrally pinned at zero both below and above the
transition (Figs.~\ref{fig:ZM_before},~\ref{fig:ZM_after}). The
winding number $\nu_3$ is a finer invariant characterizing the topological
class of that mode, the phase winding of $\phi(\tilde{\bm r})$ around the
defect, its algebraic ($1/r$) versus a more localized spatial profile, and the number of
wavefront dislocations it imprints on the local density of states. The
transition at $t'/t=2$ is a change in the latter, not the former: the
protected zero mode persists while its winding collapses, $\nu_3:\mp1\to0$.
At criticality, the symbol gap closes and $\nu_3$ is momentarily ill defined,
so the mode delocalizes and the inverse participation ratio reaches its
minimum; for $t'/t>2$, the index again guaranties a single zero mode, now
topologically trivial and exponentially localized. The order parameter of the
transition is therefore $\nu_3$, with the zero-mode count being a spectator.

For $t'/t\geq 2$, the two Dirac valleys have merged, and the valley index
is no longer an active degree of freedom.
We compute the Weyl symbol in the sublattice basis $\mathcal{B}_{AB}$.
A gradient expansion of the Wigner-Weyl transform with a vacancy at the
origin yields (derivation in Appendix~\ref{app:symbol}):
\begin{align}
H^{\text{above}}(\boldsymbol{k},\boldsymbol{r})
&= \bigl[-m - u_x k_x^2 + a^2 V_x\bigr]\sigma_x
+ \bigl[-v_y k_y + a^2 V_y\bigr]\sigma_y,
\label{eq:symbol_after}
\end{align}
where
\begin{align}
\label{eq:vacancy_above_components}
V_x &= \bigl(\tfrac{m}{2}+\tfrac{u_x}{2}k_x^2\bigr)\delta(\boldsymbol{r})
       - \tfrac{u_x}{8}\partial_x^2\delta(\boldsymbol{r})
       - \tfrac{v_y}{4}\partial_y\delta(\boldsymbol{r}), \nonumber\\
V_y &= -\tfrac{u_x}{2}k_x\partial_x\delta(\boldsymbol{r})
       + \tfrac{v_y}{2}k_y\delta(\boldsymbol{r}).
\end{align}
The operators $T=\mathbb{I}K$, $P=\sigma_z K$, $C=\sigma_z$ confirm that
the system remains in class BDI.

\paragraph*{Transition mechanism.}
Valley merging has reduced $m$ from 2 to 1.
With $d=2$ and $D=1$, condition~(\ref{eq:win_condition_chiral}) now reads
$4 \neq 2 = 2\times 1$: the dimensional constraint fails.
The winding number must vanish, $\nu_3^{\text{above}}=0$, irrespective of
vacancy perturbation details.
Physically, the emergent mass $m>0$ confines the phase of
$\tilde{f}(\boldsymbol{k})$ to a contractible arc; no non-contractible
winding can exist around the vacancy.
The transition is
\begin{equation}
    \nu_3:\; \mp 1 \;\longrightarrow\; 0 \qquad \text{at } t'/t = 2,
    \label{eq:transition}
\end{equation}
within fixed symmetry class BDI.
It is invisible in the pristine bulk and rendered observable solely through
the vacancy.

This mechanism is fundamentally different from conventional topological
phase transitions.
In the standard scenario, the bulk gap closes, the bulk invariant changes,
and edge or defect modes respond.
Here the bulk gap \emph{opens} at $t'/t=2$ while the defect invariant
simultaneously collapses.
The driving agent is the reduction in the dimension of the topological phase
space of the defect, not any spectral gap closing.
\section{Numerical results}
\label{sec:numerics}

We diagonalize the tight-binding Hamiltonian on a finite brickwall lattice
with a single type-$A$ vacancy under periodic boundary conditions.
Figure~\ref{fig:ZM_after} shows the zero-mode profile for $t'/t>2$: the mode
remains sublattice polarized, but transitions from weakly localized just above
the critical point to strongly confined deeper in the gapped phase. This is in sharp contrast to the long-ranged profiles observed for $t'/t<2$ in Figure~\ref{fig:ZM_before}.

\subsection{Inverse participation ratio}

\begin{figure}[t]
    \centering
    \includegraphics[width=0.85\linewidth]{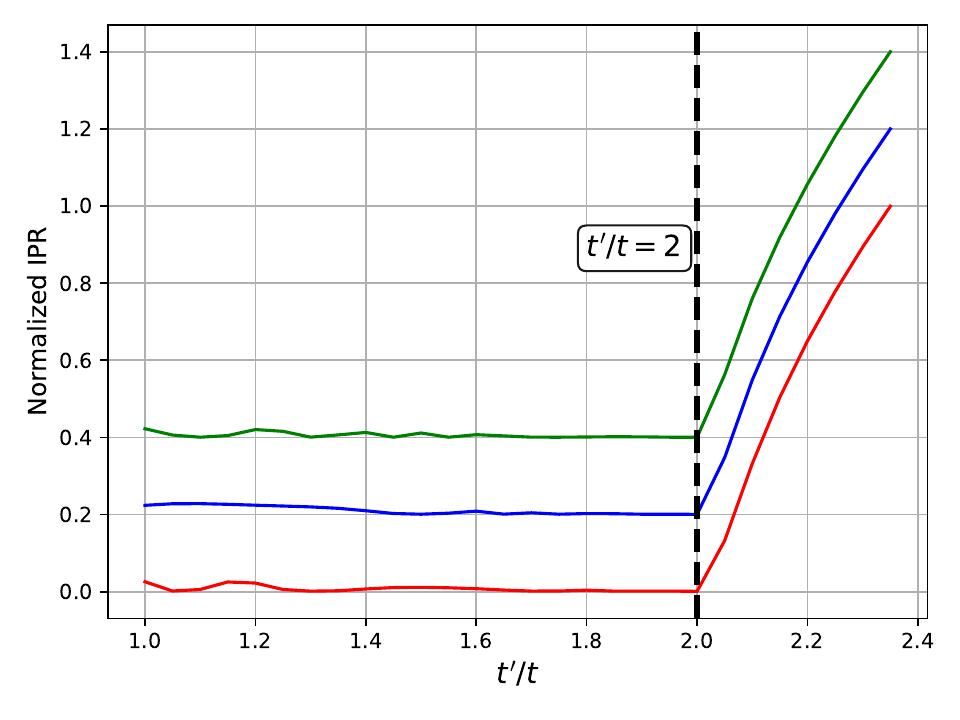}
    \caption{Normalised IPR of the vacancy zero mode vs.\ $t'/t$ for three
    system sizes ($N_x\!\times\!N_y = 160\!\times\!40$, $200\!\times\!50$,
    $248\!\times\!62$; curves offset for clarity).
    Dashed line: $t'/t=2$.
    The IPR minimum at criticality signals maximal delocalization (symbol
    gap closing); for $t'/t>2$ the IPR rises sharply.}
    \label{fig:IPR_clean}
\end{figure}

\begin{figure}[t]
    \centering
    \includegraphics[width=0.85\linewidth]{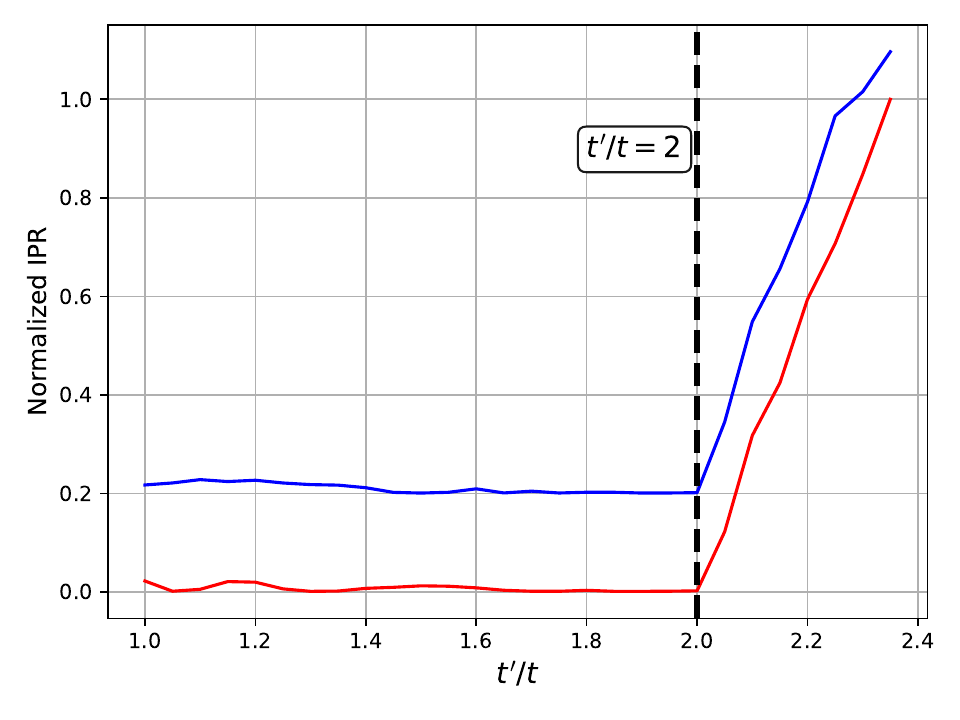}
    \caption{Disorder-averaged IPR vs.\ $t'/t$ (two system sizes;
    bond disorder $\delta t\in[-0.01t,0.01t]$; 100 realisations per point).
    The transition feature at $t'/t=2$ is robust; maximum deviation from
    the clean result is ${\sim}10\%$.}
    \label{fig:IPR_disorder}
\end{figure}

The inverse participation ratio $\mathrm{IPR}=\sum_i|\psi_i|^4$~\cite{Anderson1958}
quantifies spatial localization~\cite{Evers2000,Monthus2010}.
Figure~\ref{fig:IPR_clean} shows the IPR as a function of $t'/t$ for three
system sizes.
The curve is non-monotonic, with a pronounced minimum precisely at $t'/t=2$.
For $t'/t<2$, the mode is delocalized ($1/r$ decay) and the IPR grows slowly.
At criticality, the symbol gap closes, the Atiyah-Singer theorem no longer
applies, and the mode extends across the entire system.
For $t'/t>2$, the emergent mass gap confines the mode and the IPR rises
sharply.
This non-monotonic signature is the direct numerical fingerprint of the
topological transition.

Figure~\ref{fig:IPR_disorder} shows robustness under weak bond disorder.
Adding independent uniform perturbations $\delta t\in[-0.01t,0.01t]$ per
bond, the disorder-averaged IPR closely tracks the clean result; the maximum
deviation is ${\sim}10\%$.
The zero-energy eigenvalue remains spectrally pinned at zero under disorder,
confirming that the zero mode is topologically protected.

\subsection{Spatially resolved experimental signature}

\begin{figure}[t]
    \centering

    \begin{subfigure}{0.48\linewidth}
        \centering
        \includegraphics[width=\linewidth]{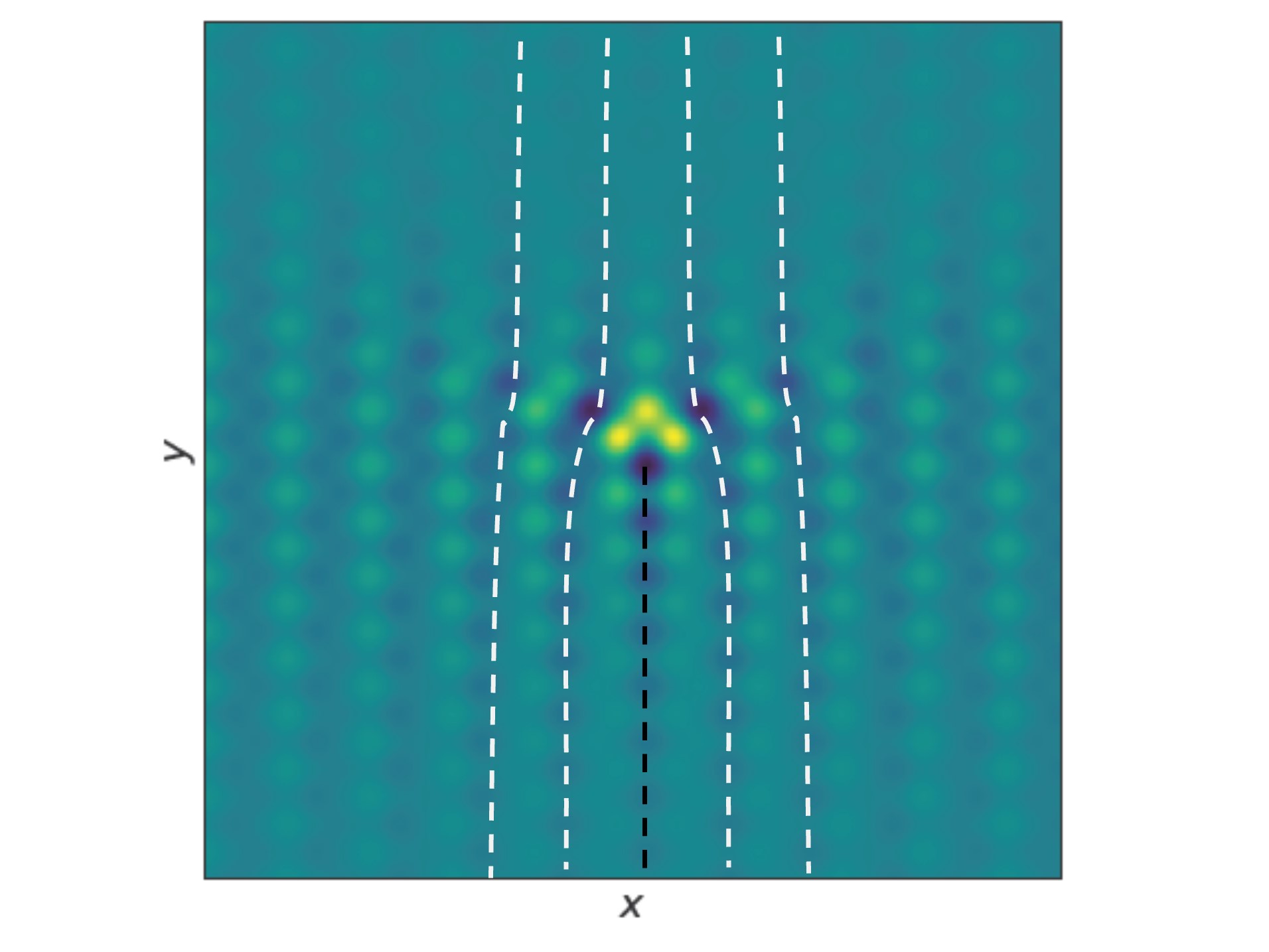}
        \caption{}
        \label{fig:Dis_tp1}
    \end{subfigure}
    \hfill
    \begin{subfigure}{0.48\linewidth}
        \centering
        \includegraphics[width=\linewidth]{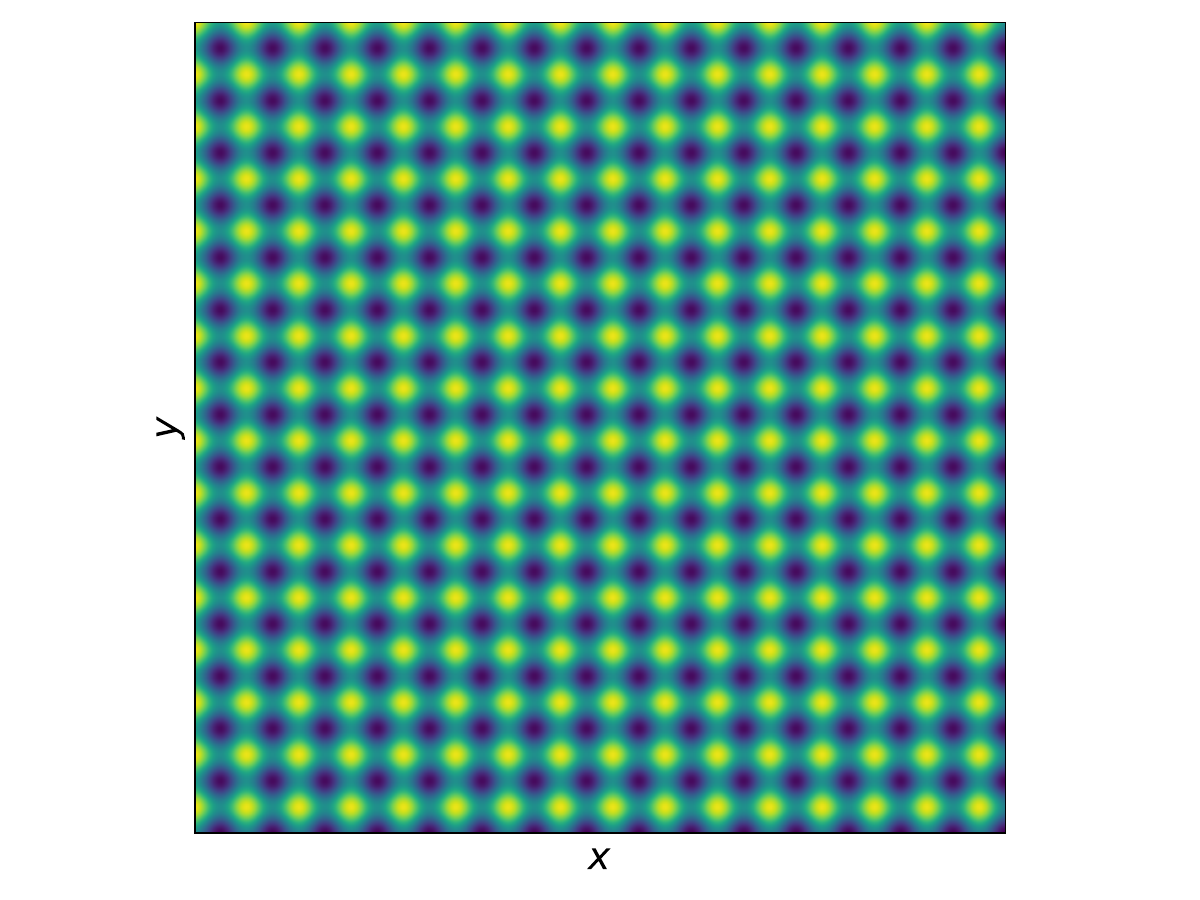}
        \caption{}
        \label{fig:Dis_tp2}
    \end{subfigure}


    \begin{subfigure}{0.48\linewidth}
        \centering
        \includegraphics[width=\linewidth]{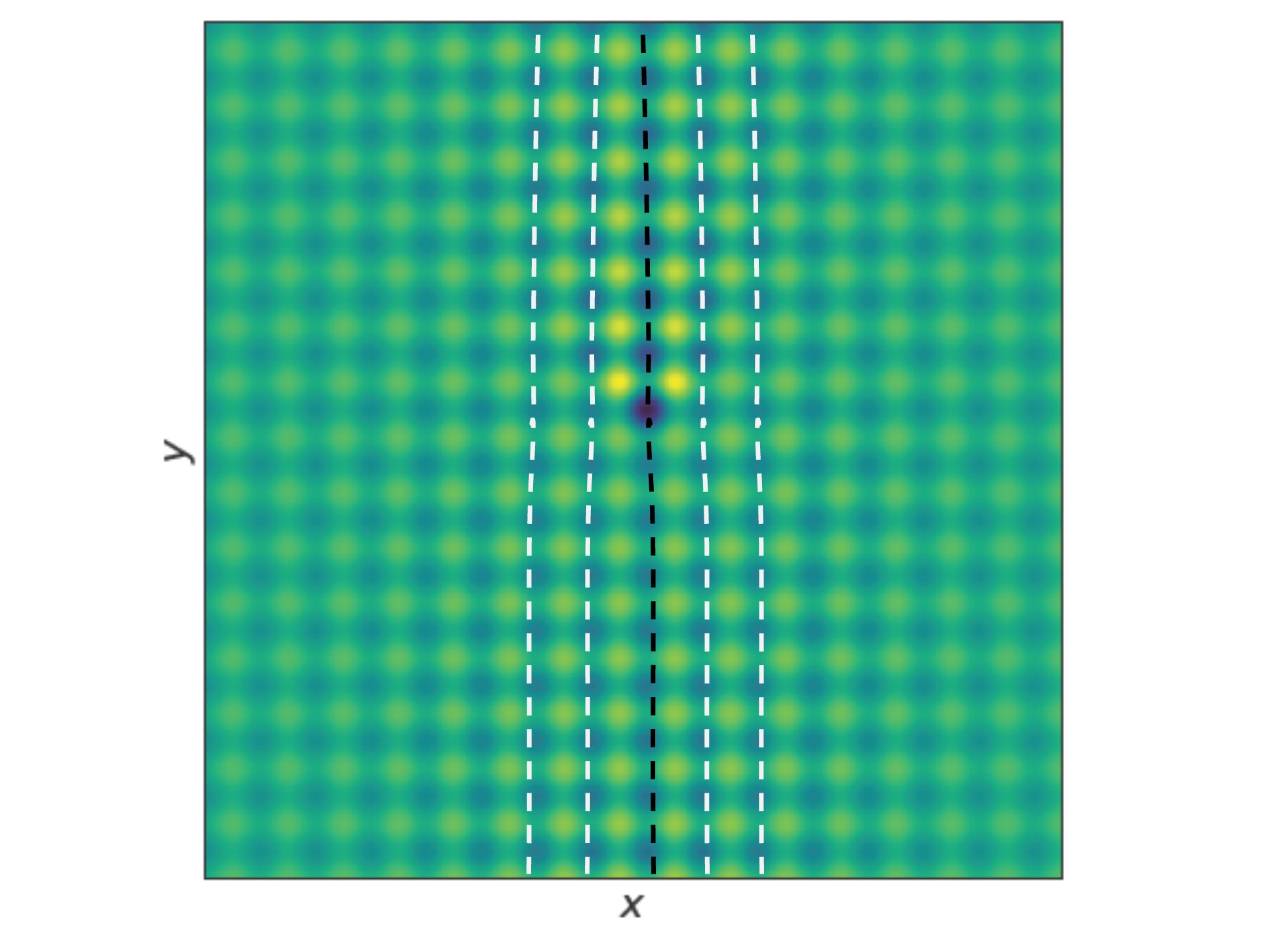}
        \caption{}
        \label{fig:Dis_tp201}
    \end{subfigure}

    \caption{Wavefront dislocations in the vacancy-induced LDOS.
    (a)~$t'/t=1$: a single dislocation, marked by the interruption of the black dashed guide line, gives $|\nu_3|=1$.
    (b)~$t'/t=2$: the dislocation disappears at the critical point.
    (c)~$t'/t=2.01$: no dislocation is present; the black guide line remains continuous, consistent with $\nu_3=0$.}
    \label{fig:dislocations}
\end{figure}


Wavefront dislocations in the LDOS induced by a vacancy encode $|\nu_3|$
directly~\cite{abulafia_wavefronts_2023}.
Figure~\ref{fig:dislocations} shows one dislocation at $t'/t=1$
($|\nu_3|=1$), its disappearance at $t'/t=2$, and its absence at
$t'/t=2.01$ ($\nu_3=0$), a spatially resolved image of
Eq.~(\ref{eq:transition}).
This is accessible by STM in graphene or by direct near-field imaging in
photonic and acoustic analogs.

\section{Discussion and Conclusions}
\label{sec:discussion}

The central result is conceptually sharp: a topological phase transition
can occur in a system whose bulk is trivial for all values of the tuning
parameter.
The transition is driven not by a bulk gap closing or symmetry breaking,
but by a change in the dimension of the topological phase space accessible
to the defect.

\paragraph*{Inversion of the bulk-boundary logic.}
In standard topological phase transitions the bulk invariant changes first,
and boundary or defect modes respond.
Here the causal direction is inverted: the bulk undergoes a Lifshitz
transition (gap opening, valley annihilation) while its invariant remains
trivially zero.
The defect carries the non-trivial winding, and the Lifshitz transition
acquires topological meaning only through it.
The vacancy acts as an internal boundary, enabling the winding that
Nielsen-Ninomiya forbids in the bulk.

\paragraph*{The $m$-counting mechanism.}
The transition is governed by condition~(\ref{eq:win_condition_chiral}).
For the honeycomb vacancy: $d=2$, $D=1$, giving $m=2$ below $t'/t=2$ (both
valleys active, $\nu_3=\mp1$) and $m=1$ above (single sublattice channel,
$\nu_3=0$).
This counting criterion predicts the critical point without detailed band
calculations, and determines the winding number in each phase.

\paragraph*{Universality.}
The criterion $d+D+1=2m$ applies to any bipartite chiral lattice where the
valley count can be continuously reduced.
Concrete candidates include:
\begin{itemize}
\item Graphene under uniaxial
    strain~\cite{Montambaux2009,Montambaux_merging2009}, where the same
    Dirac-point merging is realised mechanically.
\item Kekul\'e-distorted graphene, where the valley degree of freedom is
    controlled by the distortion amplitude.
\item Photonic and cold-atom honeycomb
    lattices~\cite{Rechtsman2013,Tarruell2012}, where hopping anisotropy is
    tunable and the LDOS dislocation count is directly imageable.
\item Three-dimensional analogs with line defects ($D=2$), where Weyl-point
    annihilation drives the transition.
\end{itemize}

\paragraph*{Outlook.}
Multiple vacancies with sublattice imbalance $|V_B-V_A|$ predict
$|\nu_3|=|V_B-V_A|$ by the index theorem; interferences between multiple
winding numbers are an open problem.
Spin-orbit-coupled variants (class DIII) may host a richer phase diagram.
The IPR minimum at criticality suggests a protocol for locating the
topological critical point experimentally.
Interaction effects near the zero mode, whose localization length diverges
at the transition, are a natural target for future work.

\section*{Acknowledgements}

This research was funded by the Israel Science Foundation Grant No.~772/21 and the Pazy Foundation.

\appendix
\renewcommand{\theequation}{A\arabic{equation}}
\setcounter{equation}{0}
\section{Derivation of the Weyl symbol above the transition}
\label{app:symbol}


For $t'/t\geq2$, we expand the tight-binding Hamiltonian in the sublattice
basis $\mathcal{B}_{AB}$ near $\boldsymbol{M}_{0}$ to obtain Eq(\ref{eq:symbol_after})-(\ref{eq:vacancy_above_components}) in the main text. The Bravais vector conventions are: $\boldsymbol{a}_{1}=a(\cos\alpha,-\sin\alpha-1)$,
$\boldsymbol{a}_{2}=a(-\cos\alpha,-\sin\alpha-1)$; nearest-neighbor
vectors: $\boldsymbol{\gamma}_{1}=a(0,1)$, $\boldsymbol{\gamma}_{2}=\boldsymbol{a}_{1}+\boldsymbol{\gamma}_{1}$,
$\boldsymbol{\gamma}_{3}=\boldsymbol{a}_{2}+\boldsymbol{\gamma}_{1}$.
The dimensionless low-energy Hamiltonian in first quantization ${\cal H}^{\text{above}}=\int d\boldsymbol{r}\psi_{\boldsymbol{r}}^{\dag}\hat{H}^{\text{above}}\psi_{\boldsymbol{r}}$:
\begin{equation}
\label{eq:H_above_decomposition}
\hat{H}^{\text{above}}
=
\hat{H}_{0}^{\text{above}}
+
\hat{V}^{\text{above}}\left(\boldsymbol{R}_{0}\right).
\end{equation}
where the pristine Hamiltonian is $\hat{H}_{0}^{\text{above}}=\begin{pmatrix}0 & -\hat{S}\\
-\hat{S}^{\dag} & 0
\end{pmatrix},$ and $\hat{S}=\left(\frac{t^{\prime}}{2t}-1-\cos^{2}\left(\alpha\right)\frac{\partial_{x}^{2}}{2}\right)-\frac{t^{\prime}}{2t}\left(1+\sin\left(\alpha\right)\right)\partial_{y}$.
For an A-sublattice vacancy located at $\boldsymbol{R}_{0}$, the
corresponding vacancy potential is $\hat{V}^{\text{above}}\left(\boldsymbol{R}_{0}\right) = a^{2} \begin{pmatrix} 0 & \delta\left(\hat{\boldsymbol{r}}-\boldsymbol{R}_{0}\right)\hat{S} \\ \hat{S}^{\dagger}\delta\left(\hat{\boldsymbol{r}}-\boldsymbol{R}_{0}\right) & 0 \end{pmatrix}.$ We compute the Wigner-Weyl transform of this operator, defined as~\cite{Case2008,Hillery1984}
\begin{equation}
    A\left(\boldsymbol{k},\boldsymbol{r}\right)=\int d\boldsymbol{r}^{\prime}e^{-i\boldsymbol{r}^{\prime}\cdot\boldsymbol{k}}\left\langle \boldsymbol{r}+\frac{\boldsymbol{r}^{\prime}}{2}\left|\hat{A}\right|\boldsymbol{r}-\frac{\boldsymbol{r}^{\prime}}{2}\right\rangle 
\end{equation}
For the pristine part, translation invariance implies that the Weyl
symbol coincides with the corresponding Bloch Hamiltonian $H_{0}^{\text{above}}\left(\boldsymbol{k}\right)=-\left(m+u_{x}k_{x}^{2}\right)\sigma_{x}-v_{y}k_{y}\sigma_{y},$ where
$\ensuremath{m=\frac{t^{\prime}}{2t}-1,u_{x}=\frac{\cos^{2}\alpha}{2},v_{y}=(\sin\alpha+1)\frac{t^{\prime}}{2t}}$.
The vacancy removes three nearest-neighbor bonds, in the Wigner-Weyl transform this introduces $\delta(\boldsymbol{r}-\boldsymbol{R}_{0})$
terms at the defect position. We consider $\boldsymbol{R}_{0}=\boldsymbol{0}$
for simplicity:
\begin{align}
\label{eq:weyl_vacancy_AB_step_1}
V_{A,B}^{\text{above}}\left(\boldsymbol{k},\boldsymbol{r}\right)
&=
a^{2}\int_{-\infty}^{\infty}d\boldsymbol{r}^{\prime}
e^{-i\boldsymbol{k}\cdot\boldsymbol{r}^{\prime}}
\left\langle
\boldsymbol{r}+\frac{\boldsymbol{r}^{\prime}}{2}
\middle|
\delta\left(\hat{\boldsymbol{r}}\right)\hat{S}
\middle|
\boldsymbol{r}-\frac{\boldsymbol{r}^{\prime}}{2}
\right\rangle
\\
\label{eq:weyl_vacancy_AB_step_2}
&=
\frac{a^{2}}{(2\pi)^2}
\int_{-\infty}^{\infty}d\boldsymbol{k}^{\prime}
e^{-i\left(\boldsymbol{k}-\boldsymbol{k}^{\prime}\right)\cdot2\boldsymbol{r}}
\left(m+u_{x}k_{x}^{\prime2}-iv_{y}k_{y}^{\prime}\right)
\\
\label{eq:weyl_vacancy_AB_step_3}
&=
\frac{a^{2}}{2}\left[
\left(m+u_{x}k_{x}^{2}-iv_{y}k_{y}\right)\delta\left(\boldsymbol{r}\right)
-iu_{x}k_{x}\partial_{x}\delta\left(\boldsymbol{r}\right)
-\frac{u_{x}}{4}\partial_{x}^{2}\delta\left(\boldsymbol{r}\right)
-\frac{v_{y}}{2}\partial_{y}\delta\left(\boldsymbol{r}\right)
\right].
\end{align}
Hermiticity gives
$V_{BA}^{\text{above}}(\boldsymbol k,\boldsymbol r)=
\left[V_{AB}^{\text{above}}(\boldsymbol k,\boldsymbol r)\right]^*.$
Combining the pristine Hamiltonian with the defect terms, we obtain Eqs.~\eqref{eq:symbol_after}--\eqref{eq:vacancy_above_components}.

\section{Reduction of the defect winding number to Eq.~(6)}
\label{app:reduction}

The defect invariant in (1) is the three–phase-space winding number
\begin{equation}
\nu_3=\frac{1}{24\pi^{2}}\oint_{S^{3}}
\mathrm{tr}\!\left[(q^{-1}dq)^{3}\right],
\label{eq:nu3-3d}
\end{equation}
where $q(\bm k,\tilde{\bm r})$ is the off-diagonal (chiral) block of the Weyl
symbol and $S^{3}$ is any surface in $(k_x,k_y,\tilde\theta)$ phase space
enclosing the defect ($d=2$ momentum directions, $D=1$ real-space angle).
We show that for the symbol~(5), this collapses to the one-dimensional phase
winding of Eq.~(6).

\paragraph{Chiral block.}
In the eigenbasis of $C=\sigma_z\otimes\mathbb I$ the symbol~(5) is
off-diagonal, $H=\mathrm{antidiag}(q,q^{\dagger})$, with the $2\times2$
($m=2$) valley block
\begin{equation}
q(\bm k,\tilde{\bm r})
=-\,i\,v_y k_y\,\mathbb I+\bm D\!\cdot\!\bm\sigma,
\qquad
\bm D=\big(\phi_1(\tilde{\bm r}),\,\phi_2(\tilde{\bm r}),\,v_x k_x\big),
\label{eq:q}
\end{equation}
the Pauli matrices are now acting in the valley space $(K,K')$.

\paragraph{Quaternionic form.}
Multiplying by the constant phase $i$ (which leaves $\nu_3$ unchanged, since
$\pi_3(U(1))=0$),
\begin{equation}
i\,q=n_0\,\mathbb I+i\,\bm n\!\cdot\!\bm\sigma,
\qquad
(n_0,\bm n)=\big(v_y k_y,\;\phi_1,\;\phi_2,\;v_x k_x\big).
\end{equation}
Thus $q$ is, up to normalization and a global phase, the unit quaternion
\begin{equation}
\hat N=\frac{\big(v_y k_y,\;\phi_1,\;\phi_2,\;v_x k_x\big)}
{\sqrt{v_x^{2}k_x^{2}+v_y^{2}k_y^{2}+\phi_1^{2}+\phi_2^{2}}}\in S^{3},
\end{equation}
and \eqref{eq:nu3-3d} equals the degree of the map
$\hat N:(k_x,k_y,\tilde\theta)\to S^{3}$. The symbol is gapped
($\hat N$ well defined) everywhere on $S^{3}$ provided
$(\bm k,\phi)\neq 0$, i.e.\ away from the band-touching point.

\paragraph{Factorisation.}
The four components of $\hat N$ separate into a momentum pair
$(v_x k_x,v_y k_y)$ and a defect pair
$(\phi_1,\phi_2)=|\phi(\tilde{\bm r})|(\cos\tilde\theta,\sin\tilde\theta)$.
$\hat N$ is therefore the join of two circle maps, and its degree
factorizes~[4,29]:
\begin{equation}
\nu_3=\mathrm{sgn}(v_x v_y)\times w[\phi],
\label{eq:factor}
\end{equation}
where $\mathrm{sgn}(v_x v_y)=\pm1$ is the chirality of the Dirac cone
(the momentum-space monopole charge) and
\begin{equation}
w[\phi]=\frac{1}{2\pi}\oint d\tilde\theta\,
\frac{\phi_1\,\partial_{\tilde\theta}\phi_2-\phi_2\,\partial_{\tilde\theta}\phi_1}
{\phi_1^{2}+\phi_2^{2}}
\label{eq:w}
\end{equation}
is the winding of the defect phase, identical to Eq.~(6). All configuration
dependence of the invariant resides in the one-dimensional integral
\eqref{eq:w}; the momentum sector contributes only the fixed sign $\pm1$.
For a single vacancy, $\phi=e^{i\tilde\theta}\phi(\tilde r)$ gives
$w[\phi]=+1$ and hence $\nu_3=\mp1$, reproducing Eq.~(6).

\paragraph{Collapse above the transition.}
For $t'/t\ge2$ the valleys merge, $v_x\to0$, and the $\sigma_z$ (valley)
component of $\bm D$ vanishes: $q$ reduces to the scalar
$-\,i\,v_y k_y+\phi\in U(1)$, i.e.\ $m=1$. Then \eqref{eq:nu3-3d} lives in
$\pi_3(U(1))=0$ and $\nu_3=0$ identically, which is the algebraic content of
condition~(1), $2m=2<d+D+1=4$.

\bibliography{Thesis_Anna}

\end{document}